# Fighting Online Click-Fraud Using Bluff Ads

Hamed Haddadi
Royal Veterinary College
University of London

## ABSTRACT

Online advertising is currently the greatest source of revenue for many Internet giants. The increased number of specialized websites and modern profiling techniques, have all contributed to an explosion of the income of ad brokers from online advertising. The single biggest threat to this growth, is however, click-fraud. Trained botnets and even individuals are hired by click-fraud specialists in order to maximize the revenue of certain users from the ads they publish on their websites, or to launch an attack between competing businesses.

In this note we wish to raise the awareness of the networking research community on potential research areas within this emerging field. As an example strategy, we present *Bluff ads*; a class of ads that join forces in order to increase the effort level for click-fraud spammers. Bluff ads are either targeted ads, with irrelevant display text, or highly relevant display text, with irrelevant targeting information. They act as a litmus test for the legitimacy of the individual clicking on the ads. Together with standard threshold-based methods, fake ads help to decrease click-fraud levels.

## Categories and Subject Descriptors

J.4 [**Computer Applications**]: Social and Behavioral Sciences—*Economics*

## 1. INTRODUCTION

Spending on Internet ads is growing faster than any other sector of the advertising industry and is expected to surge from $12.5 billion last year to $29 billion in 2010 in the U.S. alone, according to researcher eMarketer Inc. This trend is similar on the revenue form Internet activities when compared to the downfall of transit charges for example [5]. Most academics and consultants who study online advertising estimate that 10% to 15% of ads in pay per click online advertising systems are not authentic.[1] The clicks are believed to be instantiated by users in poor countries, or botnets, who are trained to click on specific ads. This process is called click-fraud.

Click fraud is the subject of some controversy and increasing litigation due to the advertising networks being a key beneficiary of the fraud. On one hand, the brokers do not wish to lose customers. On the other hand, they are not able to provide full details of the clicks, their rates, and origin IP addresses to advertisers. Even worse, the advertisers would wish to claim that all clicks are fraud, hence avoiding click charges. For instance, broker loses money to undetected click fraud when it pays out to the publisher (third party website), but it makes more money when it collects fees from the advertiser. Because of the spread between what the broker collects and what it pays out, click fraud may directly profit the broker. This also provides an incentive for the publishers to hire click fraud botnets or human teams to increase their revenue.

Brokers claim they filter out most questionable clicks and either not charge for them or reimburse advertisers that have been wrongly billed. However this is a hard task, despite the claims that brokers' efforts are satisfactory [7]. On its simplest form, a weblog publisher continuously clicks on the ads displayed on his own website in order to make revenue. In a more complicated scenario, a hardware store may hire a large, globally distributed botnet in order to click on its competitors ads, hence depleting their daily budget.

Another form of such fraud is view-fraud. In this case, a blogger may hire a botnet to continuously access its pages and retrieve ads without clicking on them, in long term generating revenue from ad impressions (views). Although the views are much cheaper than clicks, a high view to click ratio will usually affect the ad quality score and will drop the ad from the top spots, as the broker is interested in ads which generate clicks and bring in more revenue.

Detecting click fraud is a relatively new area of research. On the simplest form, the broker performs threshold-based detection. If a web page is receiving a high number of clicks from the same IP address in a short interval, these clicks can be flagged as fraud. This gets more complicated if the clickers are behind proxies or globally distributed. In such cases the broker has to use global IP blacklists and continually update its list as the botnets evolve.

In this paper we present *Bluff Ads*, a set of ads which are designed to be detected and clicked only by machines, or poorly trained click-fraud work force. These ads are targeted at the same audience profile as the other ad groups, however their displayed text is totally unrelated to the user profile. Hence they should not be clicked on by the benign user. This simple set of ads, mixed with ordinary ads, work as a litmus test, or a "Captcha" for the user legitimacy. If a high number of them is clicked, the user is deemed to be flagged as suspicious. Another form of Bluff ads is a set which contain specialized text but they are not targeted to a specific profile and are randomly displayed. This group help detect click-fraud when the botnet builds up fake profile in order to harvest relevant ads.

---
[1]http://www.businessweek.com/magazine/content/06_40/b4003001.htm

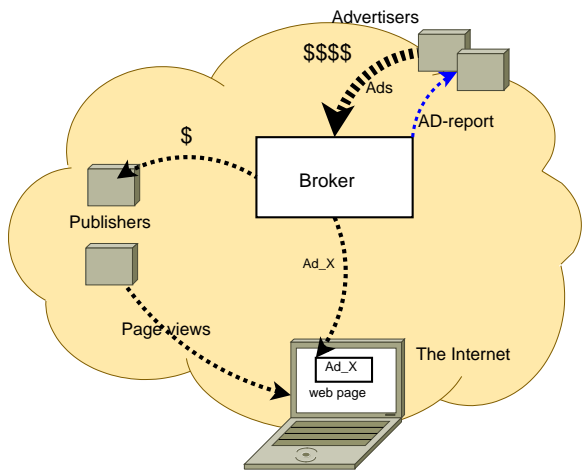

**Figure 1: Targeted keyword-based advertising.**

We wish to insist that Bluff ads are, by no means, the ultimate defense against click-fraud. In fact this paper is a report on the status quo on click-fraud and detection mechanisms which raise the bar for attackers. We wish to inspire the network security and research community as a whole to pursue further research on advertising, release more datasets (such as Yahoo! Webscope data[2]) and devise strategies for privacy preserving advertising which also minimize click fraud.

## 2. INTERNET TARGETED ADVERTISING BASICS

We define four major components of advertising systems: *advertisers*, *publishers*, *clients*, and *brokers*.

Advertisers wish to sell their products or services through ads. Publishers (e.g., personal weblogs) provider opportunities to view ads, for instance by providing space for ad banners. Clients are the computers that show publisher web pages and ads to users. Brokers (e.g., Google or Yahoo!) bring together advertisers, publishers, and clients. They provide ads to users, gather statistics about what ads were shown on which publisher pages, collect money from the advertisers, and pay the publishers.

Figure 1 illustrates the most popular advertising model on the Internet today. The advertisers specifies their ads and bids (how much the advertiser is willing to pay for views and clicks of the ads) to the broker. When a publisher provides banner space to the client on a web page, a request goes to the broker asking to fill in the banner space with an appropriate ad. The provider makes the decision as to which ads to place based on a number of criteria such as the keywords for the web page, personalization information about the client (usually persistent cookies on client machine), the keywords of the ad, and the bid associated with the ad. It then delivers the ad to the client, informs the advertiser of the ad view and clicks, and charges the advertisers and compensates the publishers accordingly.

[2]http://webscope.sandbox.yahoo.com/

## 3. CLICK FRAUD SCENARIOS

Traditionally, online advertisers used the Cost-Per-Impression (CPI) model to charge for advertisements. The CPI is often measured in terms of the cost of one thousand impressions of the ad. These ads can be videos, images and links. Recently, search engines such as Google have given rise to the Pay-Per-Click (PPC) model for online advertising. In this type of arrangement, advertisers pay a certain amount of money to the publisher for every click on their ad (which leads to the advertiser's website). This model has however led to the rise of click fraud where users, or a botnet of computers fraudulently click on the ads in order to generate revenue for the publisher in an undetectable manner. Anupam *et al.* [1] present a study on difficulty of detecting such attacks which demonstrates the difficulty of detecting click frauds. A deceitful publisher puts a script on his website that is automatically downloaded onto a visitor's computer when said visitor goes to the publisher's website. The script then imitates a click onto the advertisement leaving the visitor none the wiser. The log files of the advertiser (and, if applicable, of the publisher) will show the visitor's IP address.

Other forms of click fraud include Competitors of advertisers, vandalism fans and friends of the publisher.[3]

It is quite hard for a botnet to launch a widely distributed clicking campaign realistically. The brokers have profiles of where an advertisers' benign clickers come from, when, and how often they click. The botnet has to match this profile to avoid detection based on thresholds. The Bluff ads presented in this paper are means for detecting the nodes of a trained botnet.

Some frauds are just out of anger. In response to Google's recently announced plans to expand the tracking of users, the international anti-advertising magazine Adbusters proposes that users collectively embark on a civil disobedience campaign of intentional, automated click fraud in order to undermine Google's advertising program in order to force Google to adopt a pro-privacy corporate policy. They have released a GreaseMonkey script that automatically clicks on all AdSense ads.[4]

## 4. DETECTING CLICK FRAUD

The basic architecture for fraud detection is explained below and we will expand on different use cases in this sections.

For fraud detection, we add some Bluff adverts to the ads displayed to the user. These are real ads, but served randomly. The Bluff ads must not have a high portion of the ads. Every time the user visits a publisher page, we serve the user with probability $p(i)$ with profiled ads, and with probability $[1 - p(i)]$ with other, non related Bluff ads. The brokers' entire advertising model is based on the idea of showing only the most relevant ad content. If displayed ads are poisoned with context-free ads on a frequent basis, benign users will perceive this as the broker doing a poor job at finding relevant ads. Hence, the Bluff/real ratio must be set in a way that the user's browsing experience and ad quality perception is not greatly affected. For example, a user living in Iran should not ideally be presented with special offer ads on beer during Oktoberfest. However, it is not unreasonable, if he is shown car adverts, when his profile has no indication of his interests in driving. In practice, the Bluff ads should be authentic ads of different advertisers, spread in the network and shown randomly, but never charged for.

[3]See http://en.wikipedia.org/wiki/Click_fraud for a detailed description
[4]http://yro.slashdot.org/article.pl?sid=09/03/12/2139258&from=r

The Bluff ads serve two purposes, one, is the psychological affect, they give the user "comfort" that he is not being watched too closely and monitored too deeply, the second, and more important, purpose is to identify fraud clickers and eliminate them from the system. These fraud agents are just clicking for publishers, or against a specific advertiser.

## 4.1 Using Bluff ads

We now address different forms of click-fraud attacks and those on user privacy, and we briefly describe how the Bluff ads will help minimize them.

### 4.1.1 Profiling the customer

The Bluff ads will prevent the publishers and advertisers from narrowly targeting the client as it is not possible to know whether the ad viewed was a Bluff or not. Unless the publishers work in large groups together which will also increase the difficulty level for them. The advertisers can naturally profile users easier, however that is not avoidable as ultimately it is the users interest which lead to revenue generation for the advertiser and broker. If the broker notices that a specific web page covers many categories it can ignore that website altogether. In the case of content aggregators the broker can put specific emphasis on pages visited during clients browsing session.

### 4.1.2 Publisher fraud

The most common case is where a publisher has hired a large botnot to perform clicks for it. This can be easily realized from the frequency of the clicks. In this case the publisher ID and the Bluff/real ratio for publishers can inform the broker of this attack. If the Bluff/real ratio is higher than an average user, it is indicative of a bot being in operation.

### 4.1.3 Attacks on advertisers

These can again be identified by the a combination of simple threshold sampling and Bluff/real ratio of the ads. If most of the clicks from a host are targeted towards a single advertiser, there will be an obvious trend in their Bluff/real ratio. If the attacker decides to poison the statistics by clicking on random other ads, the ratio will be affected again. Many large advertisers today use specialized agencies to monitor their incoming traffic and identify spammers and click-fraud users, who tend to visit often but spend no time on the advertiser website. These users are also identifiable if they come from same IPs with frequent visit counts (simple threshold detection). The advertiser agents can hence pass a list of fraud suspects to the broker. They can be then removed form the billing system.

### 4.1.4 Attacks on publisher

These are the most difficult attacks, when a publisher is under attack form another source. Such attacks happen when reliable publishers who use brokers, such as CNN, are under attack in order to damage their relationships with the advertisers and providers and ultimately eject them from the competition scene. it is possible to detect such attacks by examining the Bluff/real ratio for the ad and publisher pair and identify these if the frequency of views/clicks is less than a certain time threshold. Distributed attack on publishers are a new form of attack and we have not explored the range of solutions available for detecting these in details.

## 4.2 Simple experiment

| Set | targeting age | Bluff? | Views | Clicks |
|-----|---------------|--------|-------|--------|
| 1 | 18-25 | - | 264 | 2 |
| 2 | 18-25 | YES | 140 | 0 |
| 3 | Over 25 | | 0 | 0 |
| 4 | Over 25 | YES | 0 | 0 |

Table 1: Views and clicks for different ads

In this part we present the results of a simple experiment to analyze the behavior of authentic users when presented with Bluff ads. We prepared four sets of ads using the Google Adwords Program[5], targeting UK females with keywords related to traveling to Asia. In the authentic advert texts, we mention that the travel services are specifically for female travelers between 18-25. For the bluff ads, we used the same targeting range, but used unrelated text about Internet routing in the ad text. In set 3 and set 4 of ads, we changed the targeting age, while keeping the ads the same as set 1 and 2, respectively. We ran the experiment for 3 weeks on Google Ads, on all days and hours, with even rotation and budget between the ads. We only wished to focus on the benign user reaction to Bluff ads. Hence to avoid attracting botnets we only set the advert to be displayed on Google search results pages. In this way, we avoided publisher fraud cases from biasing our ad clicks and views.

The results in Table 1 strongly support our intuition, in that when the ad text is not relevant to the audience (set 2) there is a lower number of views and clicks. Due to Google's ranking system for ad quality factor, there are less views in set 2 than set 1. set 3 and 4 represent a different set of Bluff ads, where the text of ad (in set 3) was relevant to the keywords, but the target profile (those aged 18-25) where not the ones who were targeted by the ad text. In this case the ad should have never been even viewed, in presence of a good profiling scheme, such as the one in place by Google. If the ad as such is viewed or clicked by any user, it raises the concern of that user being part of a fraud case. This simple experiment provides basic evidence on reaction of authentic users facing Bluff ads or ads which are not directly relevant to them.

## 5. RELATED WORK

There are a number of solutions for avoiding click fraud and performing better advertisement. One suggestion is to charge based on user's actions, i.e., the publisher gets a premium only after the successful conversion of the ad, meaning the user's visit to the advertiser website and performing an action such as buying an item or signing up for a service. There are a number of basic attempts at such approach by means of tracking cookies, however these efforts make up a negligible portion of the current advertising revenue on the Internet.

---

[5]https://adwords.google.com

Juels *et al.* [4] propose a cryptographic approach for replacing the pay-per-click model with one in which pay-per-action (e.g., shopping) can attract premium rates and unsuccessful clicks are discarded. in this system, the users which make a purchase are identified by the network of advertisers as premium advertisers. The client browsers use a coupon instantiated by third party cookies or issued by the attestor upon redirection. The disadvantage of this method is the ability of malicious attacker, possibly an advertiser, to use a botnet and replay the coupons numerous times, for a large number of cooperating publishers. This will then force the syndicator either discount all those replays, or removing those clients form the system with valid coupons. In both cases, the advertisement income is minimized. It also allows for the syndicator and the attestor (ad broker and middle box) to profile the users accurately including their spending budget. They also indicate that most standard click fraud techniques remain unsolved today. Despite early suggestions of this method, it has not been implemented on a large scale as it requires trust between advertisers and publishers.

Some advertisers have suggested the use of anonymised ISP data stream for verification of clicks and for better user profiling. However such attempts have been unsuccessful due to user privacy concerns. Privacy reasons also prevent brokers from releasing their server logs and click data to advertisers and their agents for deep inspection of the click rates. Other solutions include use of human-invisible ads to act as traps for botnets, however these can easily be ignored by a simple visibility test in botnet design.

Immorlica *et al.* [3] analyze the click fraud learning algorithms to compute the estimated click-through rate. They focus on a situation in which there is just one ad slot, and show that fraudulent clicks can not increase the expected payment per impression by more than o(1) in a click-based algorithm. However the complexity of the inferred algorithm and the need for click-through rate estimation would make it impractical as it also deviates form the pay per click model, to pay per view model, which is the least desired model in the modern advertisement world where bidding for space is of critical importance.

## 6. FUTURE DIRECTIONS AND CONCLUSIONS

In this paper we presented a brief overview of the current challenges in detection of click-fraud in online advertising. We presented a simple detection strategy, Bluff ads. These are sets of irrelevant ads displayed amongst user's targeted ads, which should never be clicked on. Together with threshold detection, IP address monitoring and profile matching techniques, bluff ads can be used to make it more complicated for the botnet owners to train their software, or a human operator. The bluff ads also may have a comfort factor of decreasing the user's negative perceptions by reducing the number of accurately targeted ads. We are working on deployment of bluff ads on a large network ad service. We are also hoping to design more sophisticated bluff ads in order to avoid a trained bot to easily detect their irrelevancy based on dictionary look-ups.

The current solutions to click-fraud do not detect attacks by a malicious broker. In the current advertising models, one has to fully trust the broker. There are proposals for move away from a broker-centric model [2]. However until a practical solution is deployed, research is required to minimize the impact of click-fraud on businesses.

The click-fraud issue is a new and evolving area of work, both by click-fraud investigation teams, and by attackers. We need more datasets of ad clicks, better methods of battling attacks and large scale experiments on current methods such as Bluff ads. There are click-fraud schemes in which the publisher use only real clicks, but in a way that causes the broker to pay when it should not need to do so [6]. These attacks are done by hijacking search results form `google.com` for example, and displaying results which are brought to the user by `google.com`, while in reality redirecting the search and click traffic through an attacker's website and hence claiming referral bonus from Google. In this way the user does not notice any difference in results and the conventional methods of fraud detection are hardly useful. More research from the networking community is required to understand and battle such attacks without relying on the user's spyware detection software.

## Acknowledgment

The author would like to acknowledge Christian Kreibich for insightful comments on click-fraud methods, defense mechanisms and the use of Bluff ads. I also wish to thank Richard Clegg, Jon Crowcroft, David Evans, Damien Fay, Paul Francis, David Griffin and Saikat Guha for constructive comments on earlier versions of this note.